\RequirePackage{fix-cm}
\documentclass{svjour3}                     
\smartqed  
\usepackage{graphicx}
\usepackage{amsmath}
\usepackage{amssymb}

\usepackage{mathptmx}      

\usepackage{color}

\newcommand {\be}  {
\begin{equation}
}
\newcommand {\ee}  {
\end{equation}
}
\newcommand {\bea} {
\begin{eqnarray}
}
\newcommand {\eea} {
\end{eqnarray}
}

\newcommand{\boldnabla}{\text{\boldmath$\nabla$}}

\newcommand {\ma} {
\mathcal A
}

\newcommand {\mb} {
\mathcal B
}

\newcommand {\mc} {
 \mathcal C
}

\begin{document}

\title{Elimination of surface diffusion in the non-diagonal phase field model
}


\author{Guillaume Boussinot \and
		Efim A. Brener \and
	     Claas H\"uter         \and
             Robert Spatschek
}



\institute{G. Boussinot \at
              Access e.V., RWTH 52072 Aachen, Germany, \\
              Peter-Gr\"unberg Institute, Research Center 52425 J\"ulich, Germany\\
              Tel.: +49-241-80-98013\\
              Fax: +49-241-38578\\
              \email{guillaume.boussinot@gmail.com}
           \and
            E.A. Brener \at
              Peter-Gr\"unberg Institute, Research Center 52425 J\"ulich, Germany
            \and
C. H\"uter \and R. Spatschek \at
              Computational Materials Design, Max-Planck Institute for Iron Research, 40237 D\"usseldorf, Germany 
}

\date{Received: date / Accepted: date}

\maketitle

\begin{abstract}
We present a non-diagonal phase field model for phase transformations with unequal but finite diffusivities in the two phases. This model allows to recover the desired boundary conditions at the interface, and especially the elimination of the artificial surface diffusion effect. The model is non-diagonal since it incorporates the kinetic cross coupling between the non-conserved and the conserved fields that was recently introduced [{Phys. Rev. E {\bf 86}, 060601(R), (2012)}]. We test numerically this model for the two-dimensional relaxation of a weakly perturbed interface toward its flat equilibrium.
\end{abstract}

\section{Introduction}
\label{intro}
Phase-field models are very efficient tools to tackle free boundary problems (or moving boundary problems). Especially, they have been extensively used to study growth processes (growth of a new phase at the expense of a mother one) that are coupled to a diffusional transport in the bulk \cite{wheeler,karma_rappel,antitrapping1,antitrapping2,folch}. As a time-resolved method, they are complementary to sharp-interface models, which are frequently used for steady-state simulations of microstructure evolution, see e.g. \cite{us1,us2,us3,us4}.
Phase field models describe the evolution of fields that are continuous across the interfaces between the different phases. 
One of these fields, the phase field, varies from one phase to the other, and is therefore a non-conserved quantity since one phase grows and the other shrinks. 
On the other hand, the diffusional transport in the bulk takes place because the value of a conserved quantity, for example the energy or the concentration of an alloy, is different in the two phases when they are in equilibrium with each other (at the two-phase equilibrium). Thus there is a also a conserved field in the phase field model and it obeys a continuity equation.

When the system is close enough to the two-phase equilibrium, typically when it undergoes a first-order phase transition, it is justified to use linear out-of-equilibrium thermodynamics where the driving forces are related to the fluxes through linear Onsager relations. A thermodynamically consistent procedure to derive the equations of motion within the phase field model is then provided by the so-called variational approach. In this approach, the driving forces are related to the variational derivatives with respect to the fields of a free energy or an entropy functional. 

Only very recently, a kinetic coupling between the mass flux across the interface, represented by the time derivative of the phase field, and the driving force for diffusion has been introduced in phase field models (according to Onsager reciprocity, this of course implies a kinetic coupling between the diffusional flux and the driving force for interface motion) \cite{onsager}.
This coupling between the non-conserved and the conserved fields corresponds to cross terms in the Onsager relations giving the equations of motion. The introduction of this kinetic cross coupling thus yields a non-diagonal phase field model. The importance of the new terms in the equations of evolution has been evidenced for important physical phenomena such as the solute trapping effect \cite{onsager} or the Ehrlich-Schwoebel effect \cite{mbe}. These phenomena are related to interface kinetic effects. Interface kinetics describe the deviation from the local equilibrium condition for the conserved quantity at the interface. It is negligible when the system is very close to equilibrium, for example in a Ostwald ripening regime, but may be crucial in the growth regime.

In phase field models, the magnitude of the interface kinetic effects depends strongly on the interface width, i.e. the length scale associated with the variations of the phase field. With the interface width being a numerical parameter that is artificially enhanced for computational efficiency, it is mandatory to control the interface kinetic effects in the phase field model in order to achieve the boundary conditions of the free boundary problem that one wants to reproduce. The kinetic cross coupling between the non-conserved field and the conserved field is, in this sense, of crucial importance because it allows a full correspondence between the number of degrees of freedom in the phase field model and in the macroscopic approach within which the free boundary problem is defined. A particular goal may be to recover local equilibrium boundary conditions, which are relevant for the situations, often met in experiments, where the driving force for the phase transformation is small.  

Here we are interested in achieving a control of the boundary conditions in the case where the diffusivity in the two phases is finite but differs from one phase to the other. Almgren  studied this problem in the case of thermal diffusion within a diagonal phase field model, i.e. without kinetic cross coupling \cite{almgren}. He pointed out that there exists a temperature jump at the interface. Although being a real physical effect, called the Kapitza jump, it is spuriously enhanced in phase field simulations due to the spreading of the interface, and some propositions were given in order to remove this jump. However, it was shown that completely eliminating the kinetic effects was not possible without assuming some adsorption at the interface thus altering the conservation law linking the velocity of the interface and the diffusional fluxes on both sides. Moreover, surface diffusion was shown to be a possible artificially enhanced effect that also alters the conservation law. In Ref. \cite{finite_diffusion}, the procedure to eliminate all kinetic effects without altering the conservation law was presented. However, the focus was restricted to the elimination of the Kapitza jump and only one-dimensional simulations were performed to verify the theory. The appropriate description of the conservation law was therefore not tested since this question involves two-dimensional systems. Here, we present a test of the elimination of the surface diffusion effect. For this purpose, we perform two-dimensional simulations of the relaxation of the interface to its flat equilibrium and compare the rate of the relaxation to the analytical one. The paper is organized as follows: we first briefly recall the non-diagonal phase field model for diffusional growth and then we present the numerical results.

\section{Non-diagonal phase field model}

We consider here the problem of thermal diffusion and we thus define the phase field $\phi$, that takes the value 1 in the solid and -1 in the liquid, and the temperature field $T$. We express all energies in units of the latent heat of the transformation $L=T_M(S_L-S_S)$ where $T_M$ is the melting temperature and $S_L$ ($S_S$) is the entropy of the liquid (solid) phase at $T_M$.
As mentioned in the introduction, the equations of motion for the phase field $\phi$ and the temperature field $T$ are derived within a variational approach using an entropy functional
\be
S = \int_V dV \Big\{ s[e(T,\phi), \phi] - H \left[ (1-\phi^2)^2/4 + (W \boldnabla \phi)^2/2 \right] \Big\}
\ee
where $e$ is the dimensionless internal energy density that depends on the temperature and on the phase field. For simplicity we assume that the cost of an interface is provided by the variations of the phase field only and is described by $H$. We will present later on an expression for the interfacial free energy as a function of $H$ and $W$. The internal energy obeys the continuity equation:
\be\label{continuity}
\dot e = - \boldnabla \cdot \bf J
\ee
 where $\bf J$ is the flux of dimensionless heat. The variational equations of motion relate the driving forces $\delta S/\delta \phi$ and $\boldnabla (\delta S/\delta e)$ to the fluxes $\dot \phi$ and $\bf J$ through the Onsager linear relations: 
\bea
\frac{c_P T_M^2}{L^2} \; \frac{\delta S}{\delta \phi} &=& \tau \dot \phi + (MW\boldnabla \phi) \cdot {\bf J}, \label{ds_dphi} \\
\frac{c_P T_M^2}{L^2} \; \boldnabla \frac{\delta S}{\delta e} &=&  (MW\boldnabla \phi) \dot \phi +  \frac{\bf J}{D(\phi)} \label{ds_de}, 
\eea 
where $c_P$ is the specific heat at $T=T_M$ that is assumed to be independent of the phase (independent of $\phi$). For simplicity, we consider constant $\tau$ and $M$. The thermal diffusivity $D(\phi)$ depends on $\phi$ to account for a diffusional contrast. The non-diagonal phase field model corresponds to a finite $M$, the coefficient that parametrizes the cross coupling term. We see that $\tau$, $D(\phi)$ and $M$ are the element of a symmetric matrix. This Onsager matrix has to be positive definite in order to ensure a positive entropy production, i.e. 
\be\label{stability}
\tau>0, D(\phi)>0, \Delta = 1 - (MW\boldnabla \phi)^2 D(\phi)/\tau >0.
\ee
In order to evaluate the variational derivatives $\delta S/\delta \phi$ and $\delta S/\delta e$ (see Ref. \cite{finite_diffusion} for more details), we define an internal energy density $e(T,\phi)$ and a free energy density $f(T,\phi)$ such that
\be
e(T,\phi) = u + \sigma(\phi) \;\;\; ; \;\;\; f(T,\phi) = -\frac{L}{c_P T_M}  \sigma(\phi) u
\ee 
where $u =c_P(T-T_M)/L$ and $\sigma(\phi) = (\sigma_S + \sigma_L)/2 - (\sigma_L - \sigma_S)p(\phi)/2$, with $\sigma_S = T_M S_S/L$, $\sigma_L = T_M S_L/L$ and hence $\sigma_L-\sigma_S=1$. We choose the switching function $p(\phi)= 15(\phi-2\phi^3/3+\phi^5/5)/8$ that has the properties $p(\phi = \pm 1)=\pm 1$, $p(\phi) = -p(-\phi)$ and $p'(\phi = \pm 1) = p''(\phi = \pm 1)=0$.
Then, we may present Eqs. (\ref{continuity})-(\ref{ds_de}) in the form of coupled equations of motion for $\phi$ and $u$:
\bea
\Delta \tau \dot \phi = \tilde H \left[ \phi(1-\phi^2) + W^2 \boldnabla^2 \phi \right] - \frac{p'(\phi)}{2} \; u \nonumber \\
 + M W D(\phi) \boldnabla \phi \cdot \boldnabla u  
\eea
and
\be\label{dot_u}
\dot u = \boldnabla \cdot \left \{ D(\phi) \big[ \boldnabla u + M W \dot \phi \boldnabla \phi \big]  \right\} +  \frac{p'(\phi)}{2} \; \dot \phi,
\ee
where $\tilde H = (c_P T_M^2/L^2)H$.
At equilibrium where $u=0$ and $\dot \phi=0$, the phase field obeys $\phi_{eq}(x) = - \phi_{eq}(-x)= -\tanh(x/W\sqrt{2})$ for a solid phase at $x=-\infty$ and a liquid phase at $x=+\infty$.

\section{Macroscopic boundary conditions}

 In the macroscopic description, one discusses the thermal diffusion equation:
 \be
 \dot u = D_L \boldnabla^2 u \;\; \text{ in the liquid,  } \;\; \dot u = D_S \boldnabla^2 u \;\; \text{ in the solid.}
 \ee
 where $D_L$ $(D_S)$ is the thermal diffusivity in the liquid (solid) phase. The bulk diffusion equation is supplemented with boundary conditions at the sharp interface. First, the energy conservation at the interface reads 
\bea
D_S  \boldnabla u|_S \cdot {\bf n}+ V\sigma_S =D_L  \boldnabla u|_L \cdot {\bf n}+V\sigma_L = J_E, \label{temkin}
\eea 
where $\bf n$ is the vector normal to the interface, and $\boldnabla u|_S$ ($\boldnabla u|_L$) is the gradient of $u$ on the solid (liquid) side of the interface. $V$ is the normal velocity of the interface and $J_E$ is the normal flux of dimensionless energy through the interface.
Secondly, the kinetic boundary conditions relate the difference of free energy across the interface  $\delta f$ and the Kapitza temperature jump $\delta u$ to their conjugate fluxes, respectively $V$ and $J_E$.
These relations read \cite{balibar,brener_temkin}:
\bea
(c_P T_M/L) \delta f = \sigma_S u_S - \sigma_L u_L &=& \ma V + \mb J_E+ d_0 \kappa, \label{def_kinetic1}\\
\delta u = u_L-u_S &=& \mb V + \mc J_E, \label{def_kinetic2}
\eea
where $u_S=c_P(T_S-T_M)/L$ and $u_L=c_P(T_L-T_M)/L$ with $T_S$ $(T_L)$ the temperature at the interface on the solid (liquid) side. The Gibbs-Thomson correction is proportional to the curvature $\kappa$ of the interface and is parametrized by the capillary length $d_0 =\gamma c_P T_M/L^2$. Here, the interface free energy is related to the phase field parameters through $\gamma = \alpha H T_M W$ with $\alpha = W \int_{-\infty}^{\infty} dx [\phi'_{eq}(x)]^2 = 2\sqrt{2} \big/3$ \cite{karma_rappel}. 
$\ma$ is the inverse growth kinetic coefficient, $\mc$ is the Kapitza resistance, and $\mb$ is the cross coefficient. 
 
 A usual assumption in the models for the solidification of a pure substance corresponds to 
 \be\label{bc}
 u_i=u_S=u_L= -d_0 \kappa - \ma V,
 \ee
  i.e. $\mb = \mc = 0$.
 In Ref. \cite{finite_diffusion}, it was shown using the thin interface limit that this is provided in the non-diagonal phase field model by the choices
 \be\label{one_over_d}
\frac{1}{D(\phi)} = \frac{1}{2D_S} + \frac{1}{2D_L} + g(\phi) \left(\frac{1}{2D_S} - \frac{1}{2D_L} \right)
\ee
where $g(\phi) = -g(-\phi)$ with $g(\pm 1)=\pm 1$, and
\be
M= \frac{\chi W}{2\alpha} \left( \frac{1}{2D_S} - \frac{1}{2D_L} \right).
\ee
where $\chi = W^{-1} \int_{-\infty}^\infty dx \big \{ 1-p[\phi_{eq}(x)]g[\phi_{eq}(x)] \big \}$. Then, the remaining kinetic coefficient reads
\be
\ma = \frac{\alpha \tau}{W} -\frac{\beta W }{4} \left( \frac{1}{2D_S} + \frac{1}{2D_L} \right)  \label{a}
\ee
where $\beta = W^{-1} \int_{-\infty}^\infty dx \big \{ 1-p^2[\phi_{eq}(x)] \big \} \simeq 1.40748$. The remaining degree of freedom of the phase field model to eliminate this kinetic coefficient is provided by $\tau$ that may be chosen in order to have $\ma=0$. 
In general, the full cancellation of kinetic effects may not be achieved for arbitrary $D_S/D_L$ (see Ref. \cite{finite_diffusion}) due to the conditions of stability (\ref{stability}) of the non-diagonal phase field model. The maximum contrast of diffusivity in this respect depends on the switching functions $p(\phi)$ and $g(\phi)$. In the next section where we present our numerics, we will give an explicit example for this maximum contrast. 

Almgren \cite{almgren} showed that, when $\delta u=0$, the conservation law within the phase field model reads
\bea\label{cons_pf}
V (1- \rho W \kappa) = -D_L  \boldnabla u|_L\cdot {\bf n} + D_S \boldnabla u|_S \cdot  {\bf n} - W D_{surf} \partial^2 u_i/\partial s^2
\eea
where $\rho = W^{-1} \int_{-\infty}^\infty dx \{\sigma[\phi_{eq}(x)] - \sigma_S/2 - \sigma_L/2\} = W^{-1} \int_{-\infty}^\infty dx \; p[\phi_{eq}(x)]/2$ and $D_{surf} = W^{-1}\int_{-\infty}^\infty dx [D(\phi) - D_S/2 - D_L/2]$. This conservation law deviates from the one in Eq. (\ref{temkin}) in two respects. The first concerns the factor $(1-\rho W \kappa)$ that multiplies the velocity and the deviation from unity is attributed to the interface stretching effect. There is no stretching effect with $\rho=0$ as soon as the function $p(\phi)$ is odd in $\phi$, i.e. when no adsorption excess of $\sigma[\phi_{eq}(x)]$ exists. The second is related to surface diffusion and is parametrized by $D_{surf}$. The elimination of surface diffusion, i.e. $D_{surf}=0$, may be achieved by using an appropriate function $g(\phi)$.
Again in the next section, we will present an explicit choice of $g(\phi)$ that provides this elimination.
 
 \section{Numerics}
\begin{figure}[htbp]
\begin{center}
\includegraphics[width=220pt]{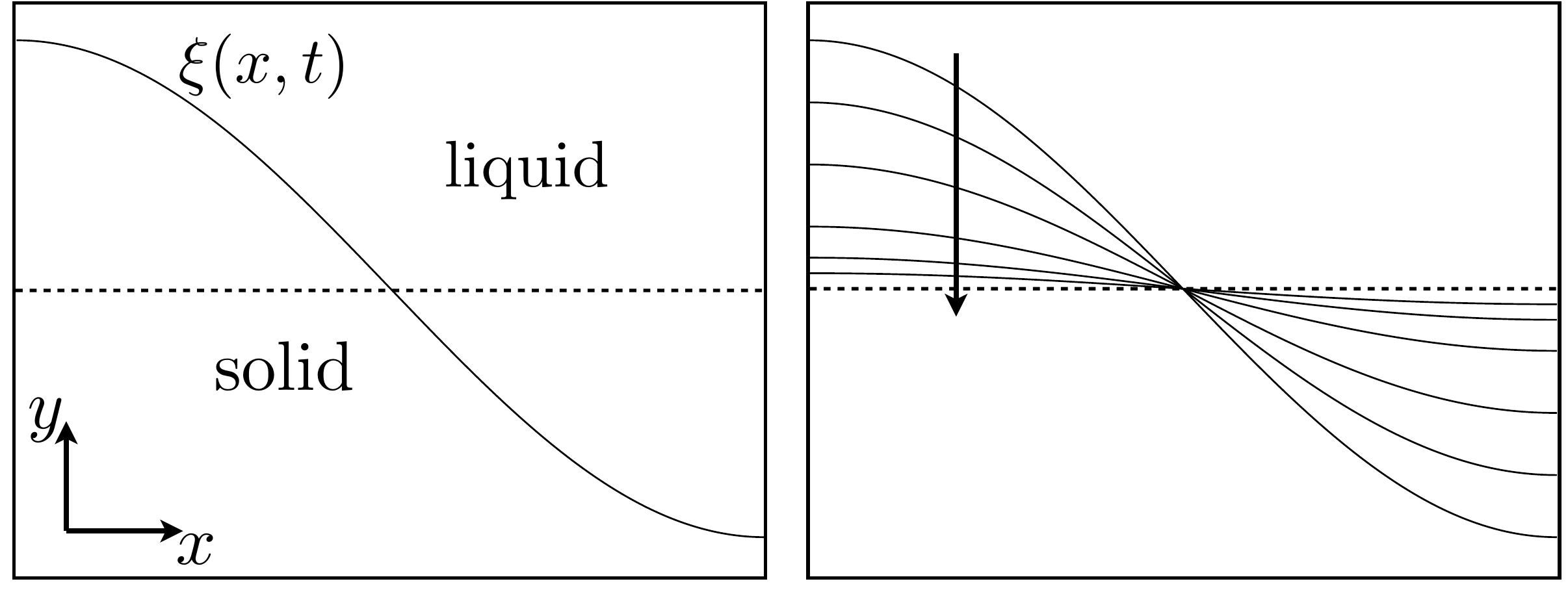}
\caption{\label{schema} Geometry of the simulated system. The solid-liquid interface is denoted by $\xi(x,t)$ and exhibits a sinusoidal perturbation. The amplitude of the perturbation decreases with time, as shown in the right panel using the vertical arrow. Here we focus on the interface but the actual size of the system in the $y$ direction is twice larger than the dimension in the $x$ direction of the half period presented here.}
\end{center}
\end{figure}

 We perform a two-dimensional numerical test of the presented theory. We study the relaxation of an interface, whose position in the $(x,y)$ plane is denoted by $y_i(x,t) = \xi(x,t)$, towards its flat equilibrium $\xi=0$ (see Fig. \ref{schema}).
Initially the interface presents a periodic perturbation $\xi(x,t=0) = \xi_0 \cos(kx)$ with $\xi_0 k \ll 1$ and the time-dependent profile of the interface is described by $\xi(x,t)= \xi_0 \cos(kx) \exp(\lambda t)$. 
The temperature field $u(x,y,t)$ obeys the diffusion equation: $\dot u = \lambda u = D_L \boldnabla^2 u$ in the liquid phase and $\dot u = \lambda u = D_S \boldnabla^2 u$ in teh solid phase. In the liquid at $y>0$, we thus have $u(x,y,t) = u_0 \exp(\lambda t + k_L y) \cos kx$ with $k_L = -\sqrt{\lambda/D_L+k^2}$ and in the solid at $y<0$, we have $u(x,y,t) = u_0 \exp(\lambda t + k_S y) \cos kx$ with $k_S = \sqrt{\lambda/D_S + k^2}$.

In the small slope approximation we have the normal velocity $V = \dot \xi(x,t) = \lambda \xi(x,t)$, the curvature $\kappa = -\partial^2 \xi(x,t)/\partial x^2 = k^2 \xi(x,t)$, the normal gradients $D_S \boldnabla u|_S \cdot {\bf n} = k_S u_i$, $D_L \boldnabla u|_L \cdot {\bf n} = k_L u_i$ and  the second derivative with respect to the arc-length $\partial^2 u_i /\partial s^2 = \partial^2 u_i /\partial x^2$, where $u_i$ is the temperature at the interface. Moreover, the boundary condition in Eq. (\ref{bc}) reads $u_i = (-d_0 k^2 - \ma \lambda) \xi(x,t)$. 
Using the conservation equation (\ref{cons_pf}), we obtain a linear equation in $\xi$ that yields:
 \bea
 \lambda = - \left( D_S \sqrt{\frac{\lambda}{D_S} + k^2} + D_L \sqrt{\frac{\lambda}{D_L} + k^2} +  W D_{surf} k^2 \right)  \left( d_0 k^2 + \ma \lambda \right).
 \eea
This rate of relaxation involves the surface diffusion effect but does not involve the stretching effect since the latter is proportional to $\xi^2$.

\begin{figure}[htbp]
\begin{center}
\includegraphics[width=180pt]{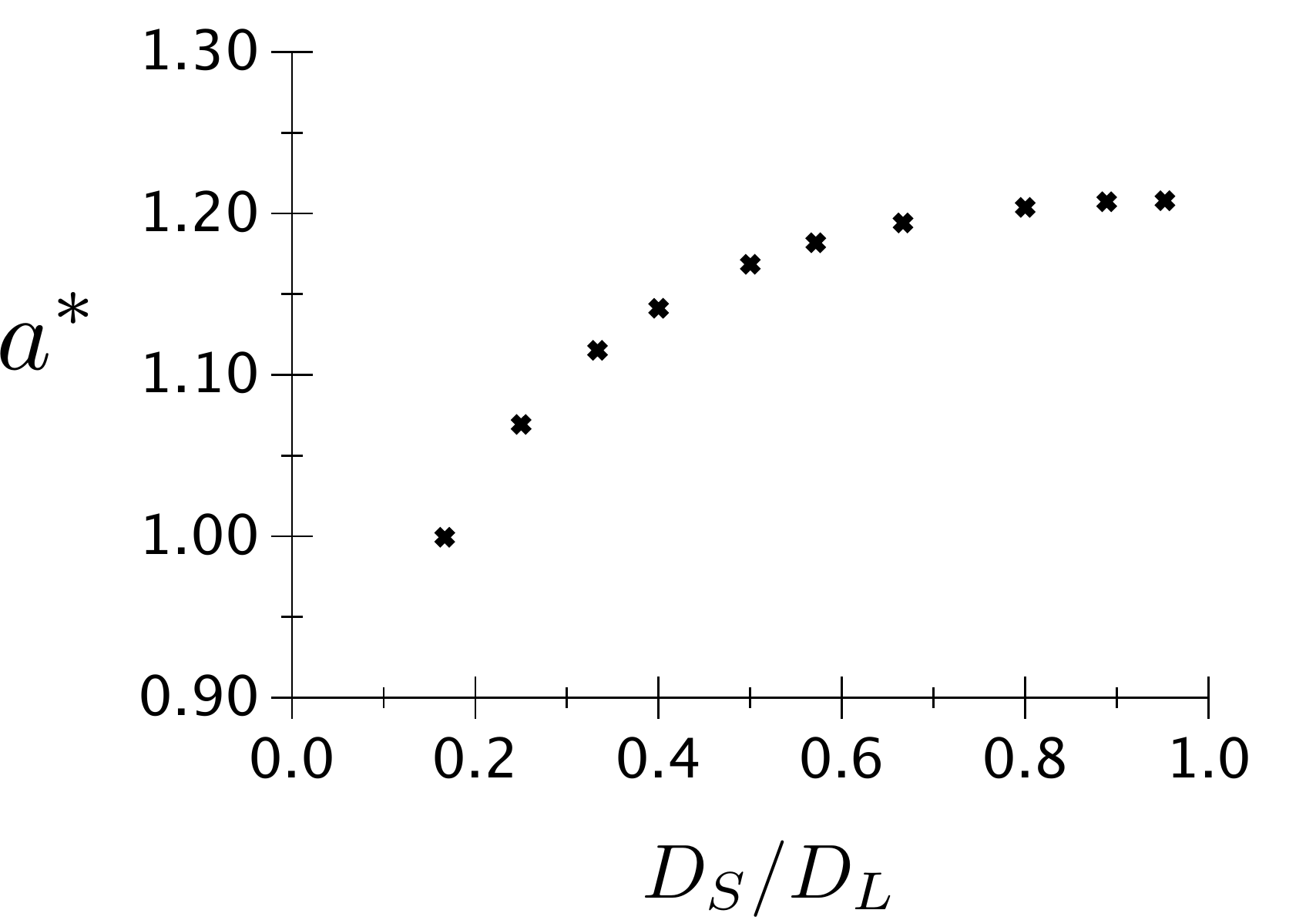}
\caption{\label{a_star} Appropriate value $a^*$ of the parameter $a$ as a function of the ratio $D_S/D_L$ for the switching functions $p(\phi)$ and $g(\phi)$ given in the text. This choice of $a$ provides an elimination of surface diffusion in the non-diagonal phase field model.}
\end{center}
\end{figure}

We did simulations with the following switching function for the diffusion coefficient [see Eq. (\ref{one_over_d})]:
\be\label{g_of_phi}
g(\phi) = \phi \left[ 1+ a(1-\phi^2) \right].
\ee
The elimination of surface diffusion $D_{surf}=0$ is provided by an appropriate choice $a=a^*$ that depends on $|(D_S-D_L)/(D_S+D_L)|$ \cite{almgren}. In Fig. \ref{a_star}, we present few values of $a^*$ as a function of $D_S/D_L$ for the choice of $g(\phi)$ given in Eq. (\ref{g_of_phi}).
As mentioned before, the condition of stability of the non-diagonal phase field model restricts the range of available $D_S/D_L$ for a full cancellation of kinetic and surface diffusion effects. Here, using $g(\phi)$ given in Eq. (\ref{g_of_phi}) and $p(\phi) = 15(\phi-2\phi^3/3+\phi^5/5)/8$, we have $1/17 < D_S/D_L < 17$. 

In the following, we present the comparison of phase field simulations of the non-diagonal model with analytics for the dimensionless rate 
\be
\tilde \lambda = -\frac{\lambda}{(D_S+D_L)d_0 k^3}.
\ee
 We use $Wk=\pi/20$ and $D_S/D_L=0.5$. We use the same discretization in both direction $\Delta x=\Delta y = 0.2 W$.

\begin{figure}[htbp]
\begin{center}
\includegraphics[width= 190pt]{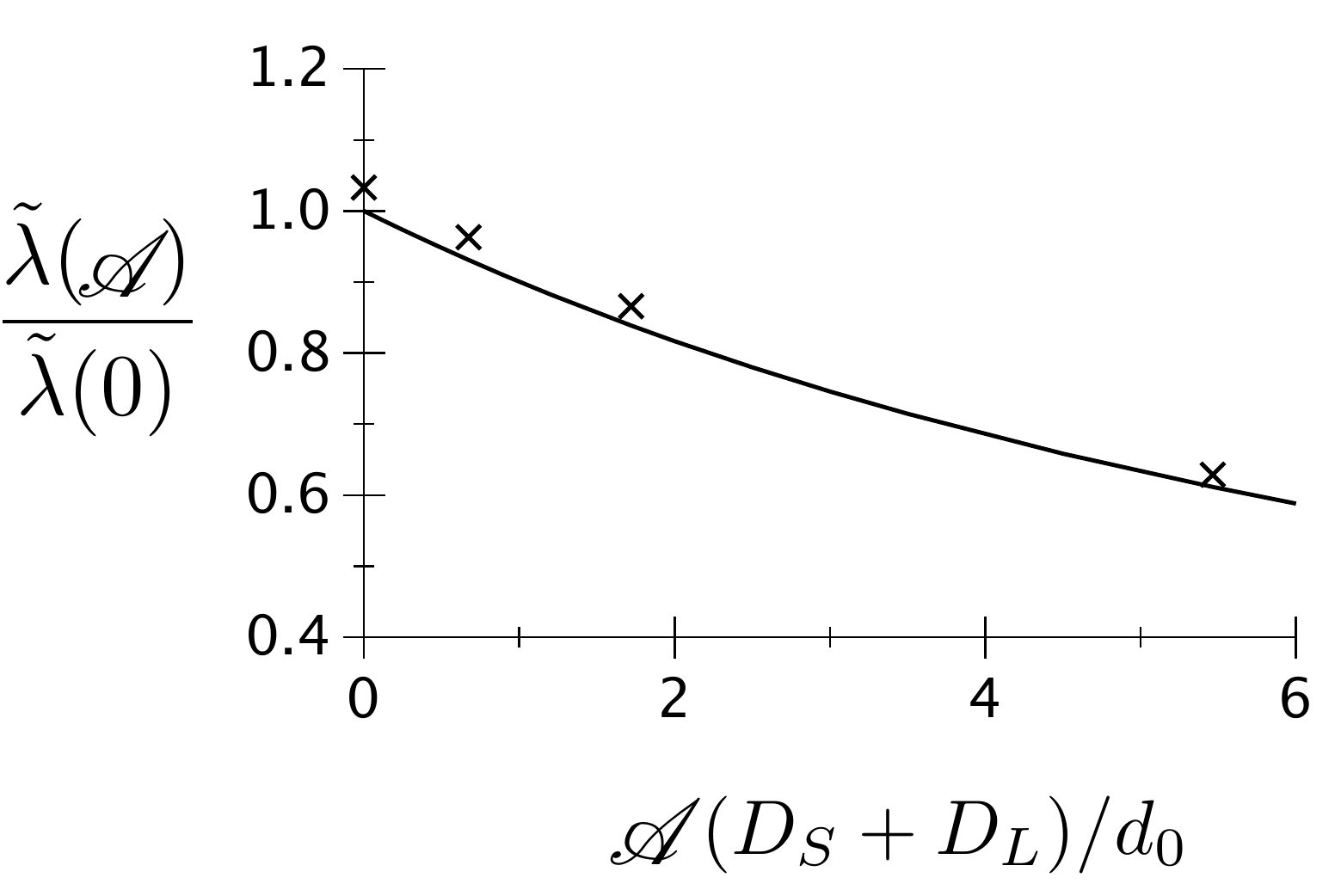}
\caption{\label{relaxation_A} Influence of the dimensionless kinetic coefficient $\ma(D_S+D_L)/d_0$ on the dimensionless rate $\tilde \lambda$ normalized by its theoretical value for $\ma=0$ (analytics: line; simulation: crosses). Here $d_0 k =\pi \sqrt{2}/30$.}
\end{center}
\end{figure}

\begin{figure}[htbp]
\begin{center}
\includegraphics[width=200pt]{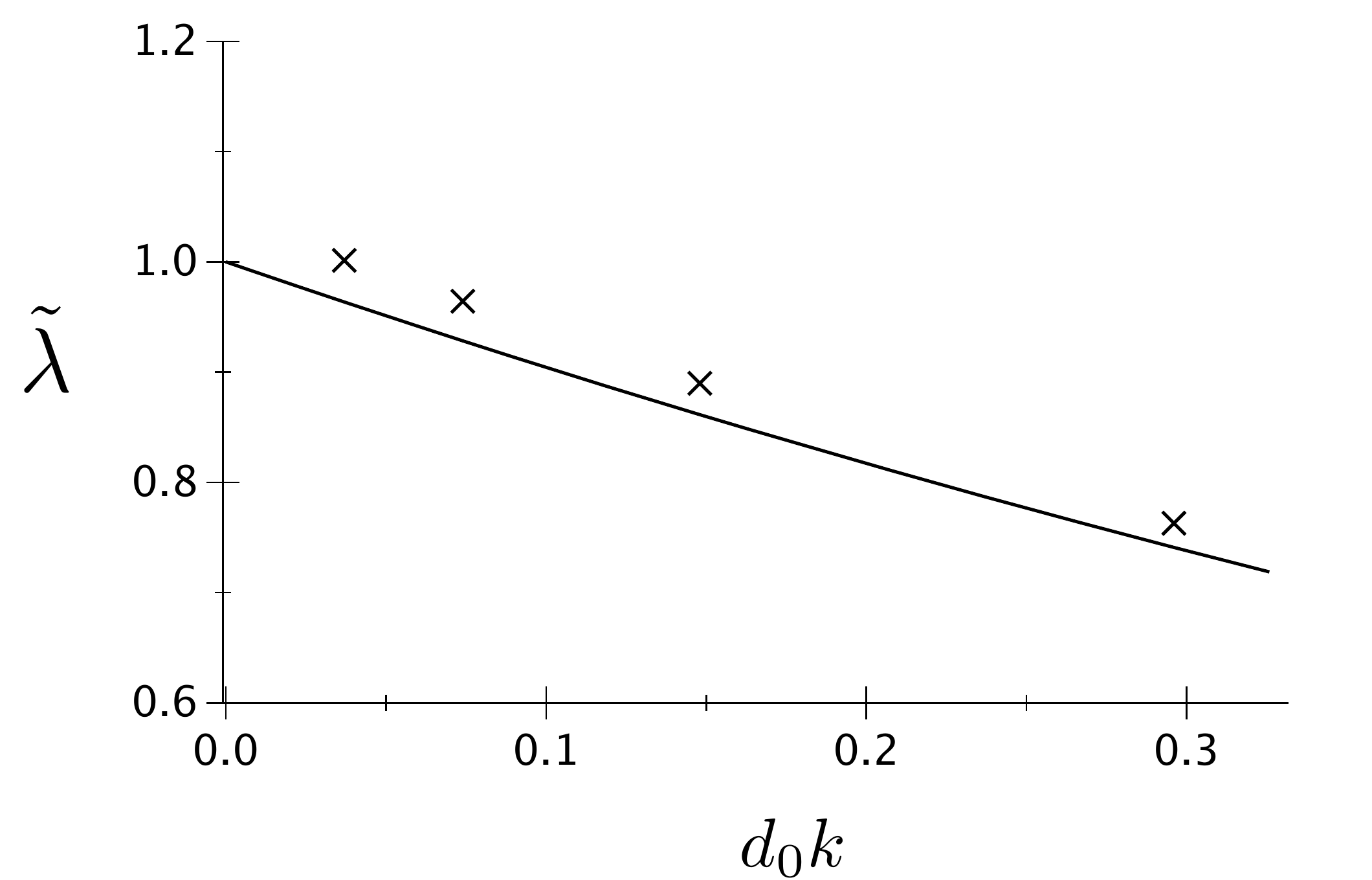}
\caption{\label{relaxation_dok} Dimensionless rate $\tilde \lambda$ as a function of $d_0 k$ for $D_{surf}=0$ and $\bar \ma=0$. The analytics are displayed with a line and the simulations with crosses.}
\end{center}
\end{figure}

In Fig. \ref{relaxation_A}, we study, with $D_{surf}=0$, the influence of the dimensionless kinetic coefficient $\ma (D_S+D_L)/d_0$ on $\tilde \lambda$ for $d_0 k = \pi \sqrt{2}/30$, i.e. for $\tilde H=1$. We normalize $\tilde \lambda$ by its theoretical value for $\ma=0$. For $D_S/D_L=0.5$, the elimination of surface diffusion $D_{surf}=0$  is provided by $a\simeq 1.1685$ (see Fig \ref{a_star}). We find a systematic 3-4\% overestimation of $\tilde \lambda$ in the simulations compared to the analytics that presumably stems for the finite discretization, and which is quite satisfactory on the quantitative level. Indeed, this accuracy should be considered in view of the deviation from the sharp interface situation quantified by the dimensionless parameter $Wk$. 
In Fig. \ref{relaxation_dok}, we present, for $D_{surf}=0$, the dimensionless rate $\tilde \lambda$ as a function of $d_0 k$. The line corresponds to the analytics and the crosses correspond to the phase field simulations. This calculations are made without interface kinetic effect, i.e. $\ma=0$. Again we obtain a systematic 3-4\% overestimation of $\tilde \lambda$ in the simulations compared to the analytics which is quite satisfactory. One should note that, as already mentioned, the conservation law in Eq. (\ref{cons_pf}) derived by Almgren \cite{almgren} requires the elimination of the Kapitza jump, i.e. $\delta u=0$, which is allowed by the introduction of the kinetic cross coupling ($M$ term). This evidences once more the importance of considering a non-diagonal phase field model.

\section{Conclusion}
In view of the results presented in Figs. \ref{relaxation_A} and \ref{relaxation_dok}, the non-diagonal phase field model reproduces well the free boundary problem where the Kapitza jump and the surface diffusion are eliminated. This agreement is provided by the recently introduced cross coupling term parametrized by $M$ and by a rather complicated input function $g(\phi)$ that has to be selected according to the ratio of diffusivities in the solid and in the liquid. The results presented in this article demonstrate that the non-diagonal model is able to achieve realistic interface kinetics and appropriate conservation laws when the diffusivity is finite in both phases but differs from one phase to the other. This issue was a long standing problem in phase field modeling of phase transformations coupled with diffusion in the bulk.

\begin{acknowledgements}

R.S. thanks the German Research Foundation (DFG) for funding via the priority program 1713.

\end{acknowledgements}




\end{document}